\documentclass[twocolumn]{aastex631}

\usepackage{graphicx}
\usepackage[none]{hyphenat}
\usepackage{amsmath}
\usepackage{booktabs}
\usepackage{multirow}
\usepackage{txfonts}
\usepackage{enumitem}
\usepackage{algorithmic}
\usepackage{bm}
\usepackage{mathrsfs}
\usepackage{booktabs}  
\usepackage{float}
\usepackage{subfigure}
\DeclareRobustCommand{\VAN}[3]{#2}
\let\VANthebibliography\thebibliography
\def\thebibliography{\DeclareRobustCommand{\VAN}[3]{##3}\VANthebibliography}
\usepackage{color}

\begin{document}

\title{Measurement of Interstellar Magnetization by Synchrotron Polarization Variance}

\author{Ning-Ning Guo}
\affiliation{Department of Physics, Xiangtan University, Xiangtan, Hunan 411105, People’s Republic of China;\\}

\author{Jian-Fu Zhang}
\affiliation{Department of Physics, Xiangtan University, Xiangtan, Hunan 411105, People’s Republic of China;\\}
\affiliation{Key Laboratory of Stars and Interstellar Medium, Xiangtan University, Xiangtan 411105, People’s Republic of China\\}
\affiliation{Department of Astronomy and Space Science, Chungnam National University, Daejeon, Republic of Korea\\}

\author{Hua-Ping Xiao}
\affiliation{Department of Physics, Xiangtan University, Xiangtan, Hunan 411105, People’s Republic of China;\\}
\affiliation{Key Laboratory of Stars and Interstellar Medium, Xiangtan University, Xiangtan 411105, People’s Republic of China\\}

\author{Jungyeon Cho}
\affiliation{Department of Astronomy and Space Science, Chungnam National University, Daejeon, Republic of Korea\\}

\author{Xue-Juan Yang}
\affiliation{Department of Physics, Xiangtan University, Xiangtan, Hunan 411105, People’s Republic of China;\\}
\affiliation{Key Laboratory of Stars and Interstellar Medium, Xiangtan University, Xiangtan 411105, People’s Republic of China\\}

\email{jfzhang@xtu.edu.cn (JFZ), hpxiao@xtu.edu.cn (HPX), jcho@cnu.ac.kr(JC), xjyang@xtu.edu.cn (XJY)}


\begin{abstract}
Since synchrotron polarization fluctuations are related to the fundamental properties of the magnetic field, we propose the polarization intensity variance to measure the Galactic interstellar medium (ISM) magnetization. We confirm the method's applicability by comparing it with the polarization angle dispersion and its reliability by measuring the underlying Alfv\'enic Mach number of MHD turbulence. With the finding of the power-law relation of $\mathcal{A} \propto M_{\rm A}^{2}$ between polarization intensity variance $\mathcal{A}$ and Alfv\'enic Mach number $M_{\rm A}$, we apply the new technique to the Canadian Galactic Plane Survey (CGPS) data, achieving Alfv\'enic Mach number of the Galactic ISM. Our results show that the low-latitude Galactic ISM is dominated by sub-Alf\'enic turbulence, with $M_{\rm A}$ approximately between 0.5 and 1.0.
\end{abstract}


\section{Introduction} 
\label{Sec_Intro}
It is generally accepted that interstellar medium (ISM) is turbulent and magnetized, and is ubiquitous in the universe (\citealt{Armstrong1995}; \citealt{Elmegreen2004}). Magnetohydrodynamic (MHD) turbulence plays a crucial role in many key astrophysical processes, such as star formation, cosmic ray acceleration and propagation, as well as magnetic reconnection (e.g., \citealt{Larson1981, Lazarian1999, Schlickeiser2003, McKee2007, Lazarian2015}). Determining the properties of magnetized turbulence is a crucial step in exploring astrophysical activity. However, the measurement of ISM magnetic field strength is extremely challenging. 

At present, several techniques measuring magnetic field strength have been proposed. As is well known, the Zeeman effect caused by spectral line splitting is a classic method in measuring the magnetic field component along the line of sight (LOS), requiring high instrument sensitivity and a longer observational integration time (\citealt{Goodman1989, Crutcher1999, Bourke2001, Crutcher2010, Mao2021, Inoue2023}). The traditional Faraday rotation synthesis can well present the magnetic structure of the Milky Way and nearby galaxies. However, this method cannot determine the magnetic field direction due to the ambiguity of the phase angle (\citealt{Burn1966, Beck2013}). The Davis-Chandrasekhar-Fermi (DCF, \citealt{Davis1951, Chandrasekhar1953}) method utilizes fluctuations in observed polarization and gas Doppler shift velocity to estimate the projected magnetic field strength by $B\propto f \sqrt{4\pi\rho}\delta v / \delta\phi$, where $f$ is an adjustable factor (typically $f=1/2$), $\delta v$ is the LOS velocity dispersion, and $\delta\phi$ is polarization angle dispersion. In general, this method results in the measured magnetic field strength deviating from the expected value. Note that this method is based on the assumption of incompressible isotropic turbulence. In fact, the theoretical and observational results imply that ISM turbulence is compressible and anisotropic (\citealt{Goldreich1995}, henceforth GS95; \citealt{Cho2002}; \citealt{Spangler2004}).

Recently, different improved versions of the DCF method have been enhanced. For instance,
\cite{Lazarian2022} proposed a differential measure approach where the adjustable factor in the DCF method is related to turbulence properties rather than a constant. In the framework of the DCF method, different effects have been included into the DCF method such as the average effect of independent eddies in the LOS direction (\citealt{Cho2016}), atomic arrangement effect (\citealt{Pavaskar2023}), and polarization angle difference as a function of the distance between a pair of measurement points (\citealt{Falceta-Goncalves2008, Hildebrand2009, Houde2009}). For the latter, due to the angle dispersion function affected by the integration effect along the LOS, the spatial range to measure should be limited within scales smaller than $0.1\ \rm pc$ (\citealt{Cho2016, Liu2021}). 

With the modern understanding of the MHD turbulence theory, other techniques for measuring the strength of magnetic fields have been proposed. First, the anisotropy of the structure functions has been proposed considering the velocity channel (\citealt{Esquivel2011, Esquivel2015}), and the velocity centroid and rotation measurement (\citealt{Xu2021, Hu2021}). Second, gradient measurements involve velocity (\citealt{Lazarian2018}) and synchrotron polarization (\citealt{Carmo2020}). Third, the ratio of the E and B modes arises from dust polarization (\citealt{Cho2023}).

In the earlier numerical work, \cite{Zhang2016} proposed the synchrotron polarization dispersion for measuring the scaling index of underlying MHD turbulence (see also \citealt{Lazarian2016} for theoretical predictions). This method is named polarization frequency analysis (PFA) which considers the variance of polarization intensity as a function of the square of wavelength $\lambda^2$. They found that the turbulent spectral index can be successfully recovered from the polarization intensity variance in the region dominated by the mean field. In this work, we want to know whether polarization intensity variance can be used as a complementary method for measuring magnetic field strength. At the same time, to enhance the reliability of our techniques, we will also explore the use of polarization angle dispersion to measure magnetization levels.\footnote{When mentioning the magnetization or magnetization levels in this paper, we specifically refer to it as the Alfv\'enic Mach number $M_{\rm A}$.} 

The structure of this paper is regularized as follows. In Section \ref{Sec_Theory}, we provide theoretical fundamentals including the properties of MHD turbulence, synchrotron polarization radiation, and statistical measure techniques. Section \ref{Sec_Data} describes simulation methods with basic parameter setup. Numerical results and application of observational data are presented in Sections \ref{Sec_Result} and \ref{Sec_observe}, respectively. Sections \ref{Sec_Discuss} and \ref{Sec_Sum} give discussion and summary, respectively.

\section{Theoretical Considerations} 
\label{Sec_Theory}
\subsection{Modern understanding of MHD Turbulence} 
\label{Subsec_MHD}
The modern understanding of the MHD turbulence theory can be traced back to the work of GS95, which focused on incompressible and trans-Alfv\'enic (with Alfv\'enic Mach number $M_{\rm A} \sim 1$) turbulence and predicted the scale-dependent anisotropy expressed as $l_{\parallel}\propto l_{\perp}^{2/3}$, where $l_{\parallel}$ and $l_{\perp}$ are the scales parallel and perpendicular to the local magnetic fields, respectively. Essentially, the turbulent cascade in the perpendicular direction and wave-like motion in the parallel direction are mutually coupled, establishing a critical balance condition by
\begin{equation}
    v_l/l_{\perp}=V_{\rm A}/l_{\parallel}
    \label{Eq1}
\end{equation}
between turbulent eddy turnover time and the Alfv\'en wave period. Here, $V_{\rm A}=B/\sqrt{\langle 4\pi\rho\rangle}$ is the Alfv\'enic velocity related to the magnetic field of $B$ and gas density of $\rho$. With this critical balance assumption, GS95 found that the cascading rate $v_l^3/l$ remains constant, i.e., $v_l\sim l^{1/3}$, which results in the classical Kolmogorov’s spectrum of $E(k)\sim k^{-5/3}$ (\citealt{Kolmogorov1941}).  

In later studies, GS95's work, in relation to trans-Alfv\'enic turbulence, was generalized to two scenarios: $M_{\rm A}=V_{\rm L}/ V_{\rm A}<1$ and $M_{\rm A} >1$ (\citealt{Lazarian1999}; \citealt{Lazarian2006}), where $V_L$ is the injection velocity at the injection or outer scale $L_{\rm inj}$. For the former, $M_{\rm A} <1$, \cite{Lazarian1999} found a transition scale 
\begin{equation}
    l_{\rm trans}=L_{\rm inj}M_{\rm A}^2.
    \label{eq_trans}
\end{equation}
In the range of $[L_{\rm inj}, l_{\rm trans}]$, with $l_{\parallel}$ remaining unchanged while $l_{\perp}$ continuously decreasing until reaching the critical equilibrium condition at the scale $l_{\rm trans}$, the turbulent cascade presents the weak turbulence properties. In the range of $[l_{\rm trans},l_{\rm diss}]$, where $l_{\rm diss}$ is the turbulent dissipation scale\footnote{In general, at the scale smaller than $l_{\rm diss}$, turbulence dissipation is dominated by numerical dissipation ($\lesssim 10$ cells, \citealt{Hernandez-Padilla2020}). If the transition scale is smaller than the dissipation scale, the numerical results connecting with the astrophysical environment may cause problems (\citealt{Brandenburg2013}).}, turbulence cascade shows the strong turbulence properties, with the parallel and perpendicular scale relationship of
\begin{equation}
    l_{\parallel}\approx L_{\rm inj}(\frac{l_{\perp}}{L_{\rm inj}})^{2/3} M_{\rm A}^{-4/3},
    \label{anisoMAless1}
\end{equation}
and with velocity and scale relationship 
\begin{equation}
    v_{\perp}=V_{\rm A}(\frac{l_{\perp}}{L_{\rm inj}})^{1/3}M_{\rm A}^{4/3}=V_{\rm L}(\frac{l_{\perp}}{L_{\rm inj}})^{1/3}M_{\rm A}^{1/3}.
\label{vel-scale-Ma}
\end{equation}
Note that when $M_{\rm A}=1$, Eqs. (\ref{anisoMAless1}) and (\ref{vel-scale-Ma}) will return to the original GS95 relationship of $l_{\parallel}\propto  l_{\perp}^{2/3}$ and $v_{\perp}\propto l_{\rm \perp}^{1/3}$, respectively.  

For the latter, $M_{\rm A}>1$, Lazarian (\citeyear{Lazarian2006}) predicted that this super-Alfv\'enic turbulence has the transition scale of
\begin{equation}
    l_{\rm A}=L_{\rm inj}M_{\rm A}^{-3},
     \label{eq_A}
\end{equation}
where the anisotropy of turbulent eddies occurs. When $M_{\rm A}>>1$, the magnetic field is not dynamically important at the maximum scale, and within the scale range of $[L_{\rm inj}, l_{\rm A}]$, turbulence follows an anisotropic Kolmogorov cascade (see Lazarian (\citeyear{Lazarian2006}) for more details).

\subsection{Synchrotron Polarization} 
\label{Subsec_SP}
When relativistic electrons move in a ubiquitous magnetic field, the synchrotron radiation process inevitably arises. With an assumption of a uniformly isotropic power-law energy distribution $N_e\propto E^{2\alpha -1}$ for relativistic electrons, where $E$ is the electron energy and  $\alpha$ is the photon spectral index related to the electron spectral index $p$ by the relation of $\alpha=(p-1)/2$, we have the synchrotron intensity $I$ per unit frequency interval $d\nu$ (\citealt{Ginzburg1965})
\begin{equation}
I(\nu) \propto \int_{0}^{L} B_{\perp}^{1-\alpha} {\nu}^{\alpha}d L^{'},
\label{intesityI}
\end{equation}
where $L$ represents the integration length along the LOS. Based on Eq.(\ref{intesityI}), we can get the linearly polarized intensity by 
\begin{equation}
P(\nu) \propto I \times \frac{3-3\alpha}{5-3\alpha}=\frac{3-3\alpha}{5-3\alpha}\int_{0}^{L} B_{\perp}^{1-\alpha} {\nu}^{\alpha}d L^{'}.
\label{intesityP}
\end{equation}
The Stokes parameters $Q$ and $U$ are related to the polarization intensity by $Q=P\cos2\phi_0$, $U=P\sin2\phi_0$, where $\phi_0$ is the polarization angle. Defining the complex polarization vector
\begin{equation}
{\boldmath{P}}=Q+iU,
\end{equation}
we can write the polarization intensity as 
\begin{equation}
P= \sqrt{Q^2+U^2},\ \phi_0=\frac{1}{2} \arctan \frac{U}{Q}.
\end{equation}

Furthermore, when synchrotron radiation experiences a radiative transfer, the Faraday rotation effect will result in the depolarization of the radiation signal. The resulting polarization angle can be re-written as
\begin{equation}
\phi=\phi_0+{\rm RM }\ {\rm \lambda}^2,
\label{Eq phi}
\end{equation}
where the rotation measure (RM) is given by 
\begin{equation}
{\rm RM}=0.81 \int_{0}^{L}n_e B_{\parallel}dz \  \rm rad\  m^{-2}.
\end{equation}
Here, $n_e$ in units of $\rm cm^{-3}$ is the thermal electron density, $B_\parallel$ in units of $\mu \rm{G}$ is the parallel component of the magnetic field along the LOS, and $dz$ in units of $\rm pc$ is the path length.

\subsection{Statistical Measure Techniques}
\label{Subsec_Var}
Since magnetic field fluctuations cause a change in polarization signal, the properties of the magnetic field can be explored through analyzing the polarization observational information. According to \cite{Zhang2016}, we define the correlation function of the polarization intensity as
\begin{equation}
\mathcal{A}=\langle P(\boldsymbol{R}_1)P(\boldsymbol{R}_2) \rangle,
\label{methodA}
\end{equation}
where $\boldsymbol{R}_1$ and $\boldsymbol{R}_2$ denote the spatial locations on the plane of the sky. Considering the correlation of polarization intensities at the same spatial separation $\Delta \boldsymbol{R} =\boldsymbol{R}_2-\boldsymbol{R}_1$=0 on the plane of the sky, we name it as the polarization variance, i.e.,
one-point statistics. Practically, we normalize the polarization intensity variance $\mathcal{A}$ in units of the mean of the polarization intensity. This procedure can eliminate the influence of dimensionality on the statistics for realistic observations in different astrophysical environments. Moreover, according to the analysis of the DCF method (e.g., \citealt{Lazarian2022}), we define the polarization angle dispersion by
\begin{equation}
\mathcal{B}=\sqrt{(\phi-\langle \phi \rangle)^2}.
\label{methodB}
\end{equation}

With the above two definitions (Eqs. (\ref{methodA}) and (\ref{methodB})), this paper will explore how $\mathcal{A}$ and $\mathcal{B}$ connect with the magnetization $M_{\rm A}$ by the Bayesian analysis method. To implement Bayesian 
statistics and fitting algorithms, we use the Python package {\it PyMC} to solve general Bayesian statistical inference and prediction problems (\citealt{Patil2010}). With the {\it PyMC}, we can determine the best fitting level and fitting error from two variables. From a theoretical view, with fundamental parameter assumptions such as the parameter $Y$ depending on the observable $X$, a linearly fitted parameter model $\beta$, and a normal distribution error $\epsilon$, we have the linear regression relationship of $Y=\beta X + \epsilon$. Therefore, this model can be represented as a probability distribution $Y\sim \mathcal{N} (\beta X,\ \sigma^2)$, where $Y$ denotes a normally distributed random variable, $\beta X$ a mean of linear predictor and a square variance of $\sigma^2$.

In practice, we consider the measured quantity $M_{\rm A}$ as the independent variable for linear fitting, and direct statistical quantities ($\mathcal{A}$ and $\mathcal{B}$) as the dependent variables. The assumed distributions and parameters of all priors in this linear model are listed as follows
\begin{equation}
    Y=\mathcal{N}(\beta M_{\rm A},\sigma^2),
\end{equation}
\begin{equation}
    \beta=\mathcal{N}(-1, 1),
\end{equation}
\begin{equation}
    \sigma^2=\lvert \rm Cauchy(0.1,2) \rvert.
\end{equation}
We use Markov Chain Monte Carlo methods in Bayesian analysis, after choosing $20, 000$ posterior \footnote{We have tested the influence of sample numbers on the results. When choosing posterior samples are greater than $10, 000$, the numerical results will remain the same.}, with each sample including $4$ parallel chains for $2, 000$ tune, the output with the highest density interval $94\%$ can provide the mean, standard deviation, and intercept of the linear fitting slope.

\section{Generation of MHD Turbulence Simulation Data} 
\label{Sec_Data}
To simulate magnetized turbulent ISM, we consider a set of governing equations of MHD turbulence as follows:
\begin{equation}
\partial\rho/\partial t + \nabla\cdot(\rho\boldsymbol{\nu})=0,
\end{equation}
\begin{equation}
\rho[\partial\boldsymbol{\nu}/\partial t + (\boldsymbol{\nu}\cdot \nabla)\boldsymbol{\nu}] + \nabla p - \boldsymbol{J}\times \boldsymbol{B}/4\pi = \boldsymbol{f},   
\end{equation}
\begin{equation}
\partial\boldsymbol{B}/\partial t -\nabla\times(\boldsymbol{\nu}\times\boldsymbol{B})=0,   
\end{equation}
\begin{equation}
\nabla\cdot\boldsymbol{B}=0,    
\end{equation}
where $\rho$ is gas density, $t$ the evolution time of the fluid, $p=c_s^2\ \rho$ gas pressure, $\boldsymbol{J}=\nabla \times \boldsymbol{B}$ the current density, and $\boldsymbol{f}$ a random driving force. At the same time, we use the isothermal state equation to close the above equations.

When numerically solving the above equations, we use a single-fluid, operator-split, and staggered-grid MHD Eulerian code ZEUS-MP/3D (\citealt{Hayes2006}). We set a non-zero uniform magnetic field $\boldsymbol{B_0}$ along the $x$-axis, resulting in a total magnetic field $\boldsymbol{B}=\boldsymbol{B_0}+\boldsymbol{\delta B}$, where $\boldsymbol{\delta B}$ is a fluctuating magnetic field, and an average density 
$\langle \rho \rangle \simeq 1$. With the periodic boundary conditions and solenoid turbulence driving, we run two sets of simulations with different resolutions: Group $A$ with a numerical resolution of 792 at a driving wavenumber of $k_{\rm inj} \simeq 2.5$, and Group $B$ with the resolution of 480 at $k_{\rm inj} \simeq 3.5$, by changing the uniform magnetic field $\boldsymbol{B}$ and injection energy. When reaching a quasi-steady state at $t\simeq 15\ t_{\rm A}$ ($t_{\rm A}$ being Alfv\'enic timescale), we terminate the simulation and then output the information of the last snapshot, which includes the 3D velocities, magnetic fields, and density. Using the definitions of Alfv\'enic Mach number $M_{\rm A}=\langle V /V_{\rm A}\rangle$, sonic Mach number of $M_{\rm s}=\langle V /c_{\rm s}\rangle$, and the plasma parameter $\beta=2M_{\rm A}^2/M_{\rm s}^2$, we characterize different data groups as listed in Table \ref{Tab_MHDdata}. The latter, $\beta$, describes the compressibility of turbulence, with a low $\beta$ corresponding to the compressible regime dominated by magnetic pressure, and a high $\beta$ to the nearly incompressible one dominated by gas pressure.

\begin{table*}
\renewcommand{\arraystretch}{1.1}
  \begin{center}
\setlength{\tabcolsep}{1.2mm}
    \begin{tabular}{cccccccccccccccccccc} 
    \hline
      \text{Models} & \text{A1} & \text{A2} &\text{A3} & \text{A4} & \text{A5} & \text{A6} & \text{A7} & \text{A8} & \text{A9} & \text{A10} & \text{B1} & \text{B2} & \text{B3} & \text{B4} & \text{B5} & \text{B6} & \text{B7} & \text{B8} & \text{B9} \\
      \hline     
      \text{$M_{\rm A}$} & 0.64 & 1.17 & 0.61 & 0.82 & 1.01 & 1.19 & 1.38 & 1.55 & 1.67 & 1.71 & 0.83 & 2.22 & 0.32 & 0.59 & 0.48 & 0.94 & 0.49 & 1.02 & 0.52\\
      \text{$M_{\rm s}$} & 0.63 & 0.60 & 5.81 & 5.66 & 5.62 & 5.63 & 5.70 & 5.56 & 5.50 & 5.39 & 0.25 & 0.24 & 0.98 & 1.92 & 0.48 & 0.93 & 0.16 & 0.34 & 0.05\\
      \text{$\beta$} & 2.06 & 7.61 & 0.02 & 0.04 & 0.07 & 0.09 & 0.12 & 0.16 & 0.18 & 0.20 & 22.74 & 172.02 & 0.21 & 0.19 & 2.02 & 2.09 & 18.99 & 17.52 & 206.56\\
      \text{$B_0$} & 0.1 & 0.03 & 4.6 & 2.6 & 1.7 & 1.1 & 0.8 & 0.6 & 0.5 & 0.4 & 0.01 & 0.001 & 1.0 & 1.0 & 0.1 & 0.1 & 0.01 & 0.01 & 0.001 \\
      \text{$\delta B_{\rm rms}/B_0$} & 0.50 & 0.86 & 0.47 & 0.63 & 0.76 & 0.87 & 1.02 & 1.12 & 1.25 & 1.39  & 0.64 & 1.84 & 0.23 & 0.46 & 0.39 & 0.74 & 0.40 & 0.80 & 0.39  \\
      \hline
    \end{tabular}
\caption{Simulation data arising from MHD turbulence. The plasma parameter $\beta=2M_{\rm A}^2/M_{\rm s}^2$ is related to $M_{\rm A}$ and $M_{\rm s}$. $B_0$ is the uniform magnetic field set initially, and $\delta B_{\rm rms}$ is the root mean square of the fluctuation field $\delta B$ as a consequence of MHD turbulence evolution. Models A1 to A10 are simulations with a numerical resolution of 792, while models B1 to B9 are simulations of 480.}
    \label{Tab_MHDdata}
  \end{center}
\end{table*}

\section{Results}
\label{Sec_Result}

\subsection{Images of polarization intensity and polarization angle with Faraday rotation}
\label{Subsec_image}

To generate synchrotron polarization observations, we use typical parameters from the Galactic ISM, such as thermal electron number density of $n_{\rm e}\sim0.01\ \rm cm^{-3}$, spectral index of relativistic electron of $p=2.5$ (note that change in the index $p$ cannot affect our results; see Zhang et al. (\citeyear{Zhang2018}) for more details), magnetic field strength of $B\sim 1.23\ {\rm \mu G}$, Faraday integral depth of $L_{\rm box}=1\ \rm kpc$.
Furthermore, we consider the $z$-axis along the LOS direction, which results in the plane of the sky parallel to the $x$-$y$ plane.
Based on Section \ref{Subsec_SP}, we obtain the images of the Stokes parameters $I$, $Q$, and $U$ using the above parameters.

\begin{figure*}[ht!]
\plotone{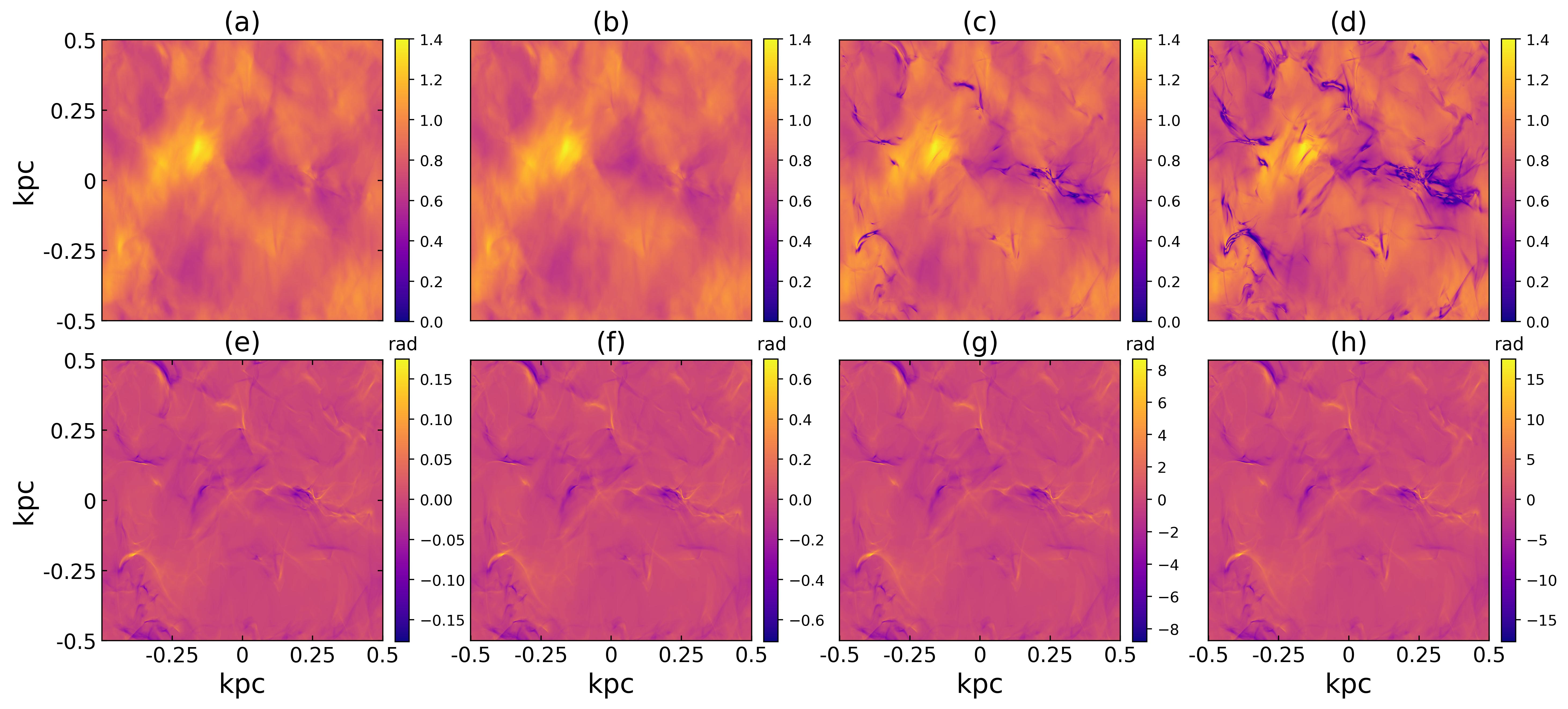}
\caption{Maps of polarization intensities (upper row) and polarization angle distributions (lower row). From left to right, the images correspond to the frequencies of 10, 5, 1.42, and 1.0 GHz, respectively. Polarization intensity is normalized in units of the mean synchrotron intensity while polarization angle is in units of a radian. The calculation is based on A3 listed in Table \ref{Tab_MHDdata}.
}
\label{Fig_image}
\end{figure*}

As is well known, when a linearly polarized electromagnetic wave propagates along a magnetic field, the polarization direction will change due to the influence of Faraday rotation. Equation (\ref{Eq phi}) indicates that the change in polarization angle is wavelength-dependent. To understand the effect of Faraday rotation on polarization radiation, we present the polarization intensity (upper row) and polarization angle (lower row) maps at different frequencies in Figure \ref{Fig_image}. From left to right, images correspond to the frequencies of 10, 5, 1.42, and $1.0\ \rm GHz$, respectively. As shown in the upper panels of Figure \ref{Fig_image}, with decreasing frequency, the maximum value of polarization intensity is slightly lower, while the minimum value is significantly lower. At the low-frequency regime (see panel (d)), we can see significantly elongated filament structures due to the enhancement of Faraday depolarization. 

The lower panels of Figure \ref{Fig_image} present the distribution of polarization angle. At the high frequency of $10\ \rm GHz$ (see panel (e)), it is $-0.18<\phi<0.18\rm\ radian$, where the Faraday rotation effect is almost negligible. As the frequency decreases, we see that Faraday rotation increases the polarization angle. When reduced to 1.42 $\rm GHz$, polarization angle rotates about $\phi\simeq \pm 3\pi\ \rm radians$ (see panel (g)).

\subsection{Finding of the power-law relationship}
\label{subsec_powerlaw}
Based on Sections \ref{Subsec_Var} and \ref{Subsec_image}, we analyze the polarization intensity variance $\mathcal{A}$ and polarization angle dispersion $\mathcal{B}$ using each MHD turbulence data listed in Table \ref{Tab_MHDdata}. Then, we use Bayesian analysis to search for the power-law relationship of $\mathcal{A}$ and $\mathcal{B}$ as a function of $M_{\rm A}$. Given the differences in the turbulence nature of MHD (see Section \ref{Subsec_MHD}), we perform a separate fitting for the sub- and super-Alfv\'enic regimes.

\begin{figure*}[ht!]
\plotone{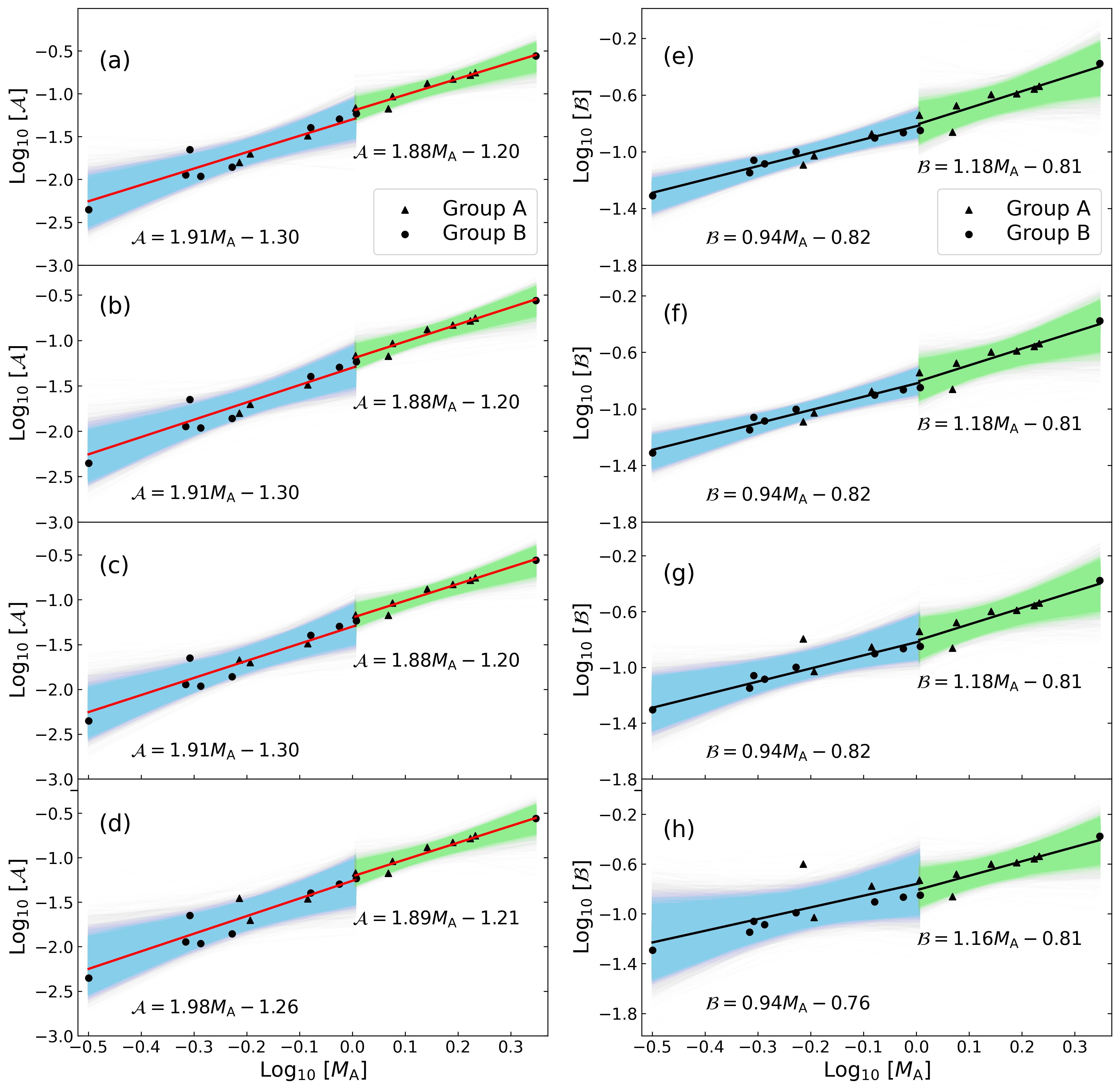}
\caption{The polarization intensity variance $\mathcal{A}$ (left column) and polarization angle dispersion $\mathcal{B}$ (right column) as a function of Alfv\'enic Mach number $M_{\rm A}$. From upper to lower, the panels correspond to 10, 5.0, 1.42, and $1.0\ \rm GHz$, respectively. 
}\label{Fig_powerlaw}
\end{figure*}

The results are plotted in Figure \ref{Fig_powerlaw} for $\mathcal{A}$ vs. $M_{\rm A}$ (left column) and $\mathcal{B}$ vs. $M_{\rm A}$ (right column). From top to bottom, the frequencies used are 10, 5, 1.42, and $1.0\ \rm GHz$, respectively. The blue and green shaded areas represent the highest density interval $94\%$, i.e., the $94\%$ fitted confidence level, in the sub- and super-Alfv\'enic regimes, respectively. In the sub-Alfv\'enic regime, we find the power-law relations of $\mathcal{A}=0.05M_{\rm A}^{1.91}$ and $\mathcal{B}=0.15M_{\rm A}^{0.94}$, while in the super-Alfv\'enic regime, we obtain the relations of $\mathcal{A}=0.06M_{\rm A}^{1.88}$ and $\mathcal{B}=0.15M_{\rm A}^{1.18}$. These relationships have been transformed from a logarithmic space to a linear space. In the range from 10 to $1.42\ \rm GHz$, we find that the power-law relation remains the same, except for the change in fitting error as summarized in Table \ref{Tab_formulae} due to the Faraday rotation effect. Although the power-law relationship between $\mathcal{A}$ and $M_{\rm A}$ changes slightly when the frequency decreases to $1.0\ \rm GHz$, we focus more on the results at a frequency of $1.42\ \rm GHz$, so this has no impact on our observational application in Section \ref{Sec_observe}.

As a result, since the polarization intensity variance and polarization angle dispersion exhibit a good power-law relationship with $M_{\rm A}$, we can employ these relations to determine the magnetization $M_{\rm A}$ of the ISM.

\begin{table}
\renewcommand{\arraystretch}{1.8}
  \begin{center}
\setlength{\tabcolsep}{2mm}
    \begin{tabular}{cccc} 
    \hline
      \text{Methods} &\text{
Frequencies} & \text{Sub-Alfv\'enic} & \text{Super-Alfv\'enic}\\
    \hline
\multirow{4}{*}{$\mathcal{A}$}  & $10\ \rm GHz$ &  $0.05^{+0.01}_{-0.02}M_{\rm A}^{1.91^{+0.59}_{-0.71}}$  & $0.06^{+0.02}_{-0.01}M_{\rm A}^{1.88^{+0.42}_{-0.38}}$  \\
& $5\ \rm GHz$ & $0.05^{+0.01}_{-0.02}M_{\rm A}^{1.91^{+0.59}_{-0.71}}$  & $0.06^{+0.02}_{-0.01}M_{\rm A}^{1.88^{+0.42}_{-0.38}}$  \\
& $1.42\ \rm GHz$ & $0.05^{+0.03}_{-0.02}M_{\rm A}^{1.91^{+0.59}_{-0.61}}$  & $0.06^{+0.02}_{-0.01}M_{\rm A}^{1.88^{+0.42}_{-0.38}}$  \\
& $1.0\ \rm GHz$ & $0.05^{+0.02}_{-0.02}M_{\rm A}^{1.98^{+0.62}_{-0.68}}$  & $0.06^{+0.02}_{-0.01}M_{\rm A}^{1.89^{+0.31}_{-0.39}}$  \\
\hline
\multirow{4}{*}{$\mathcal{B}$}  & 
$10\ \rm GHz$ & 
$0.15^{+0.02}_{-0.02}M_{\rm A}^{0.94^{+0.16}_{-0.23}}$  & $0.15^{+0.04}_{-0.03}M_{\rm A}^{1.18^{+0.42}_{-0.49}}$\\
& $5\ \rm GHz$ & $0.15^{+0.02}_{-0.02}M_{\rm A}^{0.94^{+0.16}_{-0.22}}$  & $0.15^{+0.04}_{-0.03}M_{\rm A}^{1.18^{+0.52}_{-0.47}}$\\
& $1.42\ \rm GHz$ & $0.15^{+0.06}_{-0.03}M_{\rm A}^{0.94^{+0.46}_{-0.46}}$  & $0.15^{+0.03}_{-0.03}M_{\rm A}^{1.18^{+0.52}_{-0.47}}$\\
& $1.0\ \rm GHz$ & $0.17^{+0.07}_{-0.05}M_{\rm A}^{0.94^{+0.56}_{-0.64}}$  & $0.15^{+0.04}_{-0.03}M_{\rm A}^{1.16^{+0.44}_{-0.49}}$\\
      \hline
    \end{tabular}
\caption{The power-law relationship of polarization intensity variance $\mathcal{A}$ and polarization angle dispersion $\mathcal{B}$ vs. $M_{\rm A}$ at three different frequencies.}
    \label{Tab_formulae}
  \end{center}
\end{table}

\subsection{Testing for the measurement techniques}
\begin{figure}[ht!]
\plotone{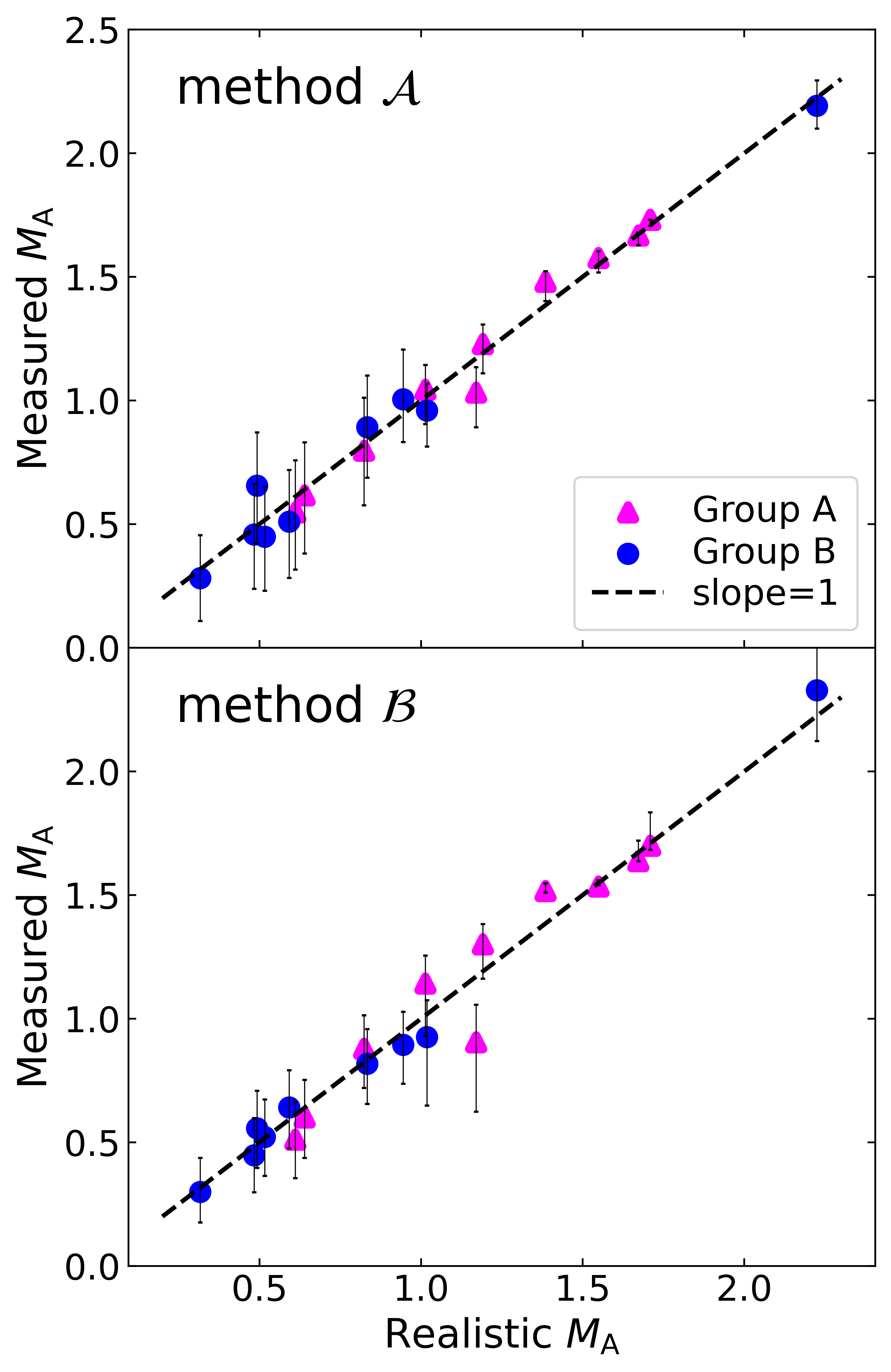}
\caption{Comparison between the measured $M_{\rm A}$ and the realistic $M_{\rm A}$. The upper and lower panels are from the polarization intensity variance and polarization angle dispersion, respectively. The errors are calculated by formulae listed in Table \ref{Tab_formulae}.
\label{Fig_comp_theor}}
\end{figure}

To test the reliability of the measurements, we first apply the two techniques to the imaging data we obtained by MHD turbulence simulations. By knowing the values of $M_{\rm A}$ in advance, we can compare their values with the $M_{\rm A}$ measured by techniques. Figure \ref{Fig_comp_theor} shows the results we obtained by polarization intensity variance (upper panel) and the polarization angle dispersion (lower panel). The error bars are calculated using the upper and lower error limits summarized in Table \ref{Tab_formulae}, which also indicates the range of fitting errors presented in Figure \ref{Fig_powerlaw}. From Figure \ref{Fig_comp_theor}, we can see that filled data points for circles and triangles are distributed near the dashed line with a slope equal to 1, which demonstrates that the $M_{\rm A}$ measured by both methods is consistent with the realistic $M_{\rm A}$ determined from 3D MHD turbulence simulations. 

Consequently, the power-law relationships we obtained are reliable for measuring $M_{\rm A}$. Next, we apply the power-law relationships summarized in Table \ref{Tab_formulae} to measure the magnetization of the Galactic ISM.  

\section{Application to Observations} 
\label{Sec_observe}
Here, we apply our techniques to the Canadian Galactic Plane Survey (CGPS) data with arcminute-scale images of all major components of the ISM over a large portion of the Galactic disk. Note that the Dominion Radio Astrophysical Observatory (DRAO) Synthesis Telescope surveys have imaged a $73$ section of the Galactic plane. The DRAO fields were combined into $1024\times 1024$ pixel mosaic images ($5 \fdg12\times5\fdg12$ for each), with a resolution close to $1^\prime$ on a Galactic Cartesian grid at the 1.42 GHz. By randomly selecting 9 mosaics named MA1, MF2, MST2, MU1, MU2, MW1, MEV1, MEV2, and MEX1 (see \citealt{Taylor2003} for detailed coordinate locations), we extract $800\times 800$ pixels, corresponding to the region of $4\fdg0 \times 4\fdg0$, from individual mosaic images to avoid the margin effect of the images (see also \citealt{Zhang2019}).

We consider a statistical average of different sub-regions from the whole mosaic image to measure expected $M_{\rm A}$ from the whole mosaic image. In practice, we first decompose the selected $4 \fdg0\times4\fdg0$ mosaic image into 400 and 256 sub-block regions, which correspond to $0 \fdg20\times0\fdg20$ and $0 \fdg25\times0\fdg25$ pixels, respectively. Then, we employ the power-law relationships summarized in Table \ref{Tab_formulae} to obtain the $M_{\rm A}$ values of each sub-block region. Finally, we obtain statistical measurements of $M_{\rm A}$ for all sub-regions to characterize the magnetization of the whole mosaic image.

\begin{figure*}[ht!]
\plotone{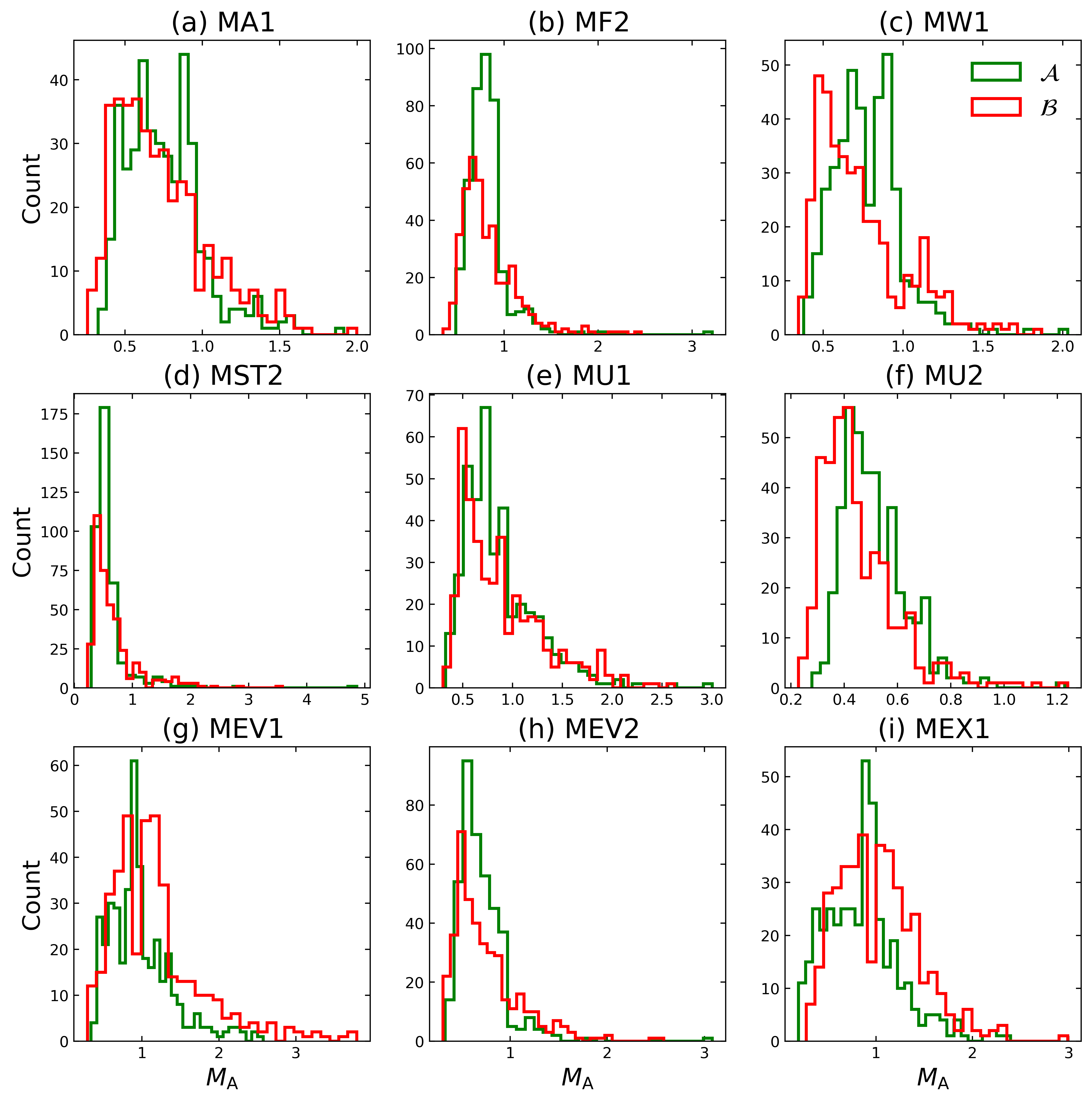}
\caption{The measured $M_{\rm A}$ distribution of CGPS mosaic images with the size of $0 \fdg20\times0\fdg20$ pixels. The red and green histograms correspond to the distribution of $M_{\rm A}$ calculated by the polarization intensity variance and polarization angle dispersion, respectively.}
\label{Fig_obs_40size}
\end{figure*}

The $M_{\rm A}$ measurement results using $0 \fdg20\times0\fdg20$ pixels are plotted in Figures \ref{Fig_obs_40size}, from which we can see that histogram distributions from methods $\mathcal{A}$ and $\mathcal{B}$ are approximately consistent. To obtain quantitative $M_{\rm A}$ values, we characterize the measurement for each mosaic image by the average and peak values of histogram distributions. Fitting the $M_{\rm A}$ distribution histogram with the Gaussian function at a $95\%$ confidence level, we obtained the average and peak values of the $M_{\rm A}$ distribution for each mosaic image. 

\begin{figure*}[ht!]
\plotone{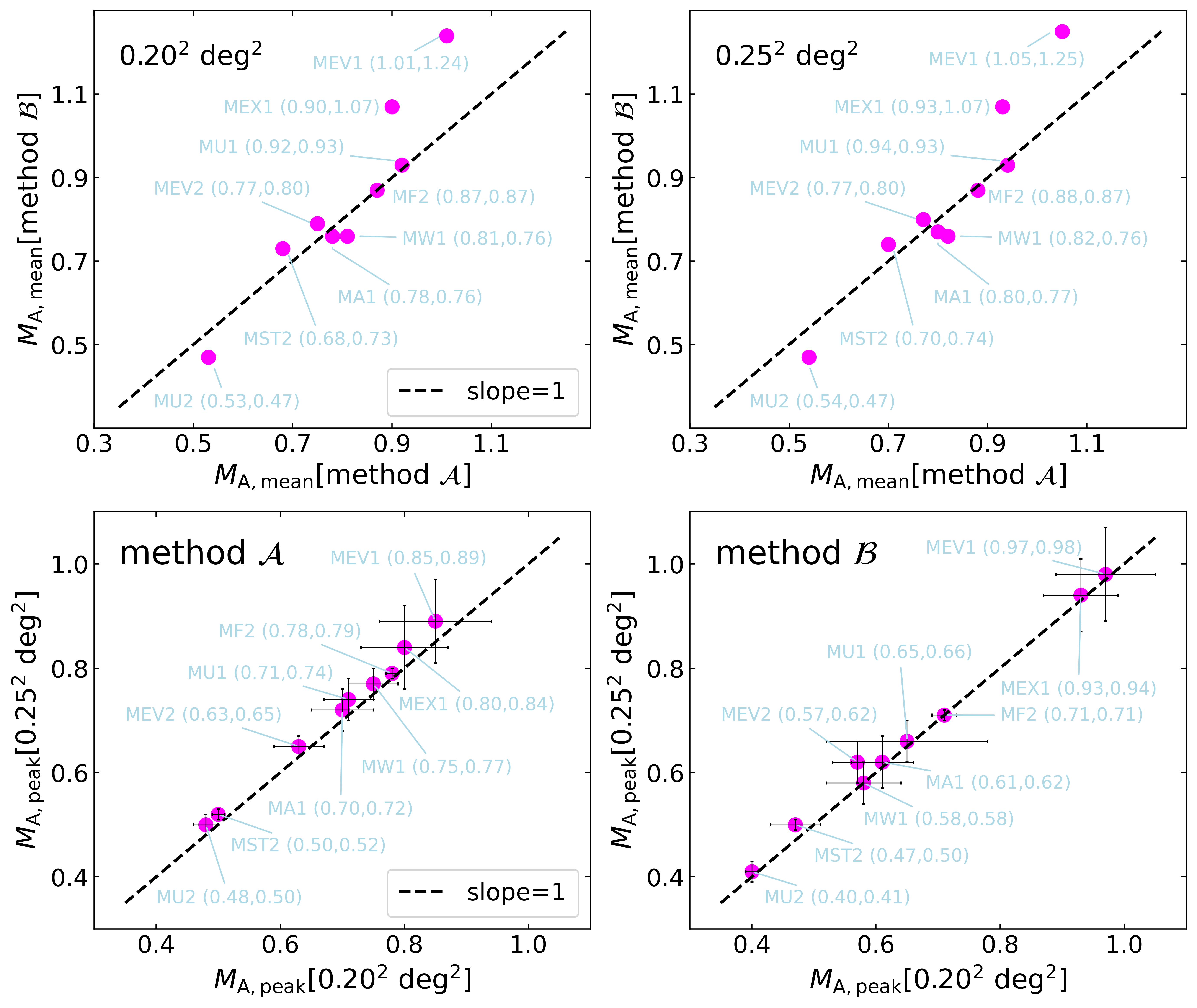}
\caption{Upper panels: Comparison of the average values of $M_{\rm A}$ distribution between methods $\mathcal{A}$ and $\mathcal{B}$ when the sub-region size is $0 \fdg20\times0\fdg20$ (left upper) and $0 \fdg25\times0\fdg25$ (right upper) pixels, respectively. Lower panels: Comparison of the $M_{\rm A}$ distribution peak values between sub-region sizes of $0 \fdg20\times0\fdg20$ and $0 \fdg25\times0\fdg25$ pixels, arising from methods $\mathcal{A}$ (left lower) and $\mathcal{B}$ (right lower).}
\label{Fig_obs_com}
\end{figure*}

The upper panels of Figure \ref{Fig_obs_com} show the comparison of the average $M_{\rm A}$ values measured by methods $\mathcal{A}$ and $\mathcal{B}$ when selecting different sub-region sizes. As seen in the upper panels, the results measured by the two methods are relatively similar in most regions, except that the results of method $\mathcal{B}$ are about 0.1 larger for the MEV1 and MEX1 mosaics. This result shows that on the one hand, the measurement results of method $\mathcal{A}$ are more reliable, and on the other hand, the size of the selected sub-region has no effect on measurements featured by average values. The lower panels of Figure \ref{Fig_obs_com} compare the $M_{\rm A}$ distribution peak values of the same method when changing the size of the sub-region sizes, with the error bar plotted by a Gaussian function fitting error. It can be seen that the measurement results of methods $\mathcal{A}$ and $\mathcal{B}$ are the same in different sub-region sizes, indicating that the choice of sub-region size has little effect on the measurement results of $M_{\rm A}$. As a result, we propose that the low-latitude Galactic ISM is mainly dominated by sub-Alfv\'enic turbulence, with a magnetization level approximately between 0.5 and 1.0.

\section{Discussion}
\label{Sec_Discuss}
Our current work has explored how polarization intensity variance can be used to measure the ISM magnetization, confirming the feasibility of this method. Considering the influence of Faraday depolarization on the measurement method, we find that in the explored parameter space, even if the polarization angle rotation reaches $2\pi$ radians (see Figure \ref{Fig_image}), the power-law relation we established is less affected by the Faraday rotation effect, that is, only the fitting error variation of Bayesian statistics (see Table \ref{Tab_formulae}).  

To further confirm the complementarity of the polarization intensity variance measurements, we have also tested other methods such as the top-base ratio from polarization gradient statistics, inverted variance method, and polarization degree dispersion. These methods have been explored in \cite{Lazarian2018} and \cite{Carmo2020}. Our testing results are similar to Figure 7 of \cite{Lazarian2018} using the Galactic Arecibo L-band Feed Array $\rm HI$ data. Note that the measurement of the polarization degree dispersion often provides a smaller measurement value, because the Faraday rotation effect will result in a relatively low observed polarization degree. 

We would like to emphasize that our current work is based on the framework of the modern understanding of MHD turbulence theory. The established power-law relationship is confined to the strong turbulence regime of the MHD turbulence cascade (see Section \ref{Subsec_MHD}), that is, we ensure that the MHD simulation dataset listed in Table \ref{Tab_MHDdata} has a sufficiently wide inertial range, with $l_{\rm trans}$ and $l_{\rm A}$ greater than $l_{\rm diss}$ (see Equations (\ref{eq_trans}) and (\ref{eq_A})). For polarization angle dispersion, our results arising from compressible turbulence show the power-law relation of $\mathcal{B} \propto M_{\rm A}^{0.9}$ and $\mathcal{B} \propto M_{\rm A}^{1.2}$ in the sub- and super-Alfv\'enic regimes, respectively. Differently, \cite{Skalidis2021} and \cite{Lazarian2023} claimed that the polarization angle dispersion maintains the relation of $\mathcal{B} \propto M_{\rm A}$ for incompressible turbulence, and $\mathcal{B} \propto M_{\rm A}^2$ for compressible turbulence. What we would like to clarify is that their polarization dispersion is defined as $\delta \phi = \delta B_{\perp}/B$, which is different from our definition in Equation (\ref{methodB}). Note that \cite{Falceta-Goncalves2008} mentioned that the polarization angle dispersion may be related to the angle between the LOS direction and the mean magnetic field. In our work, we do not consider the influence of the angle between the LOS and the mean magnetic field on our new method; this will be explored in the future. 

In general, when the brightness temperature approaches the thermal temperature of relativistic electrons, the synchrotron self-absorption effect is expected to be important at low frequencies \citep{Zhang2022}. However, our testing shows that this radiative transfer process is negligible for the parameters used in this paper. In addition, it is not important for the effects of nonadiabatic propagation for Faraday rotation in the Galactic ISM \citep{Lee2019}.

\section{Summary}
\label{Sec_Sum}
In this work, we proposed a new method, polarization intensity variance, for measuring ISM magnetization. After confirming the reliability of the new technique using numerical simulation data, we applied the technique to measure the Galactic ISM magnetization with real observations. The main results are summarized as follows.

\begin{itemize}
\item We proposed that polarization intensity variance can achieve accurate measurement of the ISM magnetization. Compared with the polarization angle dispersion method, we verify the reliability of the polarization intensity variance method in measuring the ISM magnetization.

\item We found a power-law relationship of $\mathcal{A} \propto M_{\rm A}^{2}$ between polarization intensity variance $\mathcal{A}$ and magnetization $M_{\rm A}$. For the polarization angle dispersion, the power-law relationships are $\mathcal{B} \propto M_{\rm A}^{0.9}$ in the sub-Aflv\'enic turbulence and $\mathcal{B} \propto M_{\rm A}^{1.2}$ in the super-Aflv\'enic turbulence.

\item Using simulation polarization data with the known $M_{\rm A}$ values, we demonstrated that polarization intensity variance can successfully recover the underlying magnetization $M_{\rm A}$, supporting the applicability of this new method.

\item With the application of polarization intensity variance, we propose that the low-latitude Galactic ISM is dominated by sub-Alf\'enic turbulence, with $M_{\rm A}$ approximately between 0.5 and 1.0.

\end{itemize}

\begin{acknowledgments}
We would like to thank the anonymous referee for the valuable comments that improved our manuscript. J.F.Z. thanks for the support from the National Natural Science Foundation of China (grant No. 11973035), the Hunan Natural Science Foundation for Distinguished Young Scholars (No. 2023JJ10039), and the China Scholarship Council for the overseas research fund. H.P.X. thanks the support from the Scientific Research Foundation of the Education Bureau of Hunan Province (No. 23A0132). X.J.Y is supported in part by NSFC 12333005 and 12122302 and CMS-CSST-2021-A09.
\end{acknowledgments}

\bibliography{sample631}{}
\bibliographystyle{aasjournal}

\end{document}